\documentclass{PoS}

\title{Precise Determinations of the Decay Constants of $B$ and $D$ mesons}

\ShortTitle{Decay Constants of $B$ and $D$ mesons}

\author{\speaker{Heechang Na} $^a$, Chris Monahan$^b$, Christine Davies$^c$, Eduardo Follana$^d$, Ron Horgan$^e$, Peter Lepage$^f$, Junko Shigemitsu$^g$\\
        \llap{$^a$}ALCF, Argonne National Laboratory, Argonne, IL 60439, USA\\
        	\llap{$^b$}Department of Physics, College of William and Mary, Williamsburg, VA 23187, USA\\
	\llap{$^c$}SUPA, School of Physics \& Astronomy, University of Glasgow, Glasgow, G12 8QQ, UK\\
	\llap{$^d$}Departamento de Fisica Teorica, Universidad de Zaragoza, E-50009 Zaragoza, Spain\\
	\llap{$^e$}DAMTP, Cambridge University, Cambridge, CB3 0WA, UK\\
	\llap{$^f$}LEPP, Cornell University, Ithaca, NY 14853, USA\\
	\llap{$^g$}Department of Physcis, The Ohio State University, Columbus, OH 43210, USA\\
        E-mail: \email{heena@alcf.anl.gov}}

\abstract{Recently we studied the $B$, $B_s$, $D$ and $D_s$ meson decay constants using various treatments for the heavy quark. 
For $B$ mesons, we determined $f_B$, $f_{B_s}$, and $f_{B_s}/f_B$ with NRQCD bottom quarks.
We then combined the ratio $f_{B_s}/f_B$ and another very precise determination from HPQCD for $f_{B_s}$ using heavy HISQ quarks, and extracted $f_B$ with 2\% total errors. We also calculated $f_D$, $f_{D_s}$, and $f_{D_s}/f_D$ using HISQ charm quarks.
Here we review our results and briefly discuss their implications for the determination of the CKM matrix elements $|V_{cd}|$ and $|V_{cs}|$.
}

\FullConference{The 30th International Symposium on Lattice Field Theory - Lattice 2012,\\
		June 24-29, 2012\\
		Cairns, Australia}

\begin{document}

\section{Introduction}
Investigating the flavor structure of the Standard Model (SM) is important for its own sake; however, it is even more interesting since it can lead to physics beyond the SM.
Furthermore, the data accumulations and new analysis emerging from the LHC suggest more and more that the Higgs is very close to the SM Higgs. 
They even do not find any hint of new particles yet. 
In this situation, precise understanding of the SM in the flavor sector becomes more critical.  

Decay constants of heavy-light pseudoscalar mesons have been studied from lattice QCD for quite some time. 
We can determine the corresponding CKM matrix elements by combining decay constants from theory and decay rates from experiments. 
Moreover, decay constants are basic quantities related to many hadronic quantities.
For instance, $f_B$ is an important input parameter for inclusive determinations of the CKM matrix elements. 
The decay constants can also be used for testing the lattice formalism, since the calculations have typically smaller errors and the procedures are relatively straightforward.
  
This paper presents new calculations for $B$, $B_s$, $D$ and $D_s$ meson decay constants from HPQCD. We have completed the projects, and published the results in two papers~\cite{fb}\cite{fd}.
So, essentially this proceeding consists of a brief summary of the two papers and a discussion of their impact on the CKM matrix elements $|V_{cd}|$ and $|V_{cs}|$.


\section{$B$ and $B_s$ meson decay constants}
We used the MILC AsqTad $N_f=2+1$ gauge configurations with NRQCD (Nonrelativistic QCD) $b$ quarks for this project.
The previous HPQCD calculation~\cite{fb2} used AsqTad light and strange valence quarks; however, in this work we used the highly improved staggered quark (HISQ) action for the valence quarks. 
The HISQ action has much smaller discretization effects, so one can expect improvements in the continuum extrapolation errors.
We include one more ensemble (F0); $40^3\times96$ with $m_l/m_s=0.0031/0.031$, which is a more-chiral fine ensemble. Details of the lattice configurations are in Tab.~\ref{enb}. 

\begin{figure}
\includegraphics[width=.45\textwidth,angle=270]{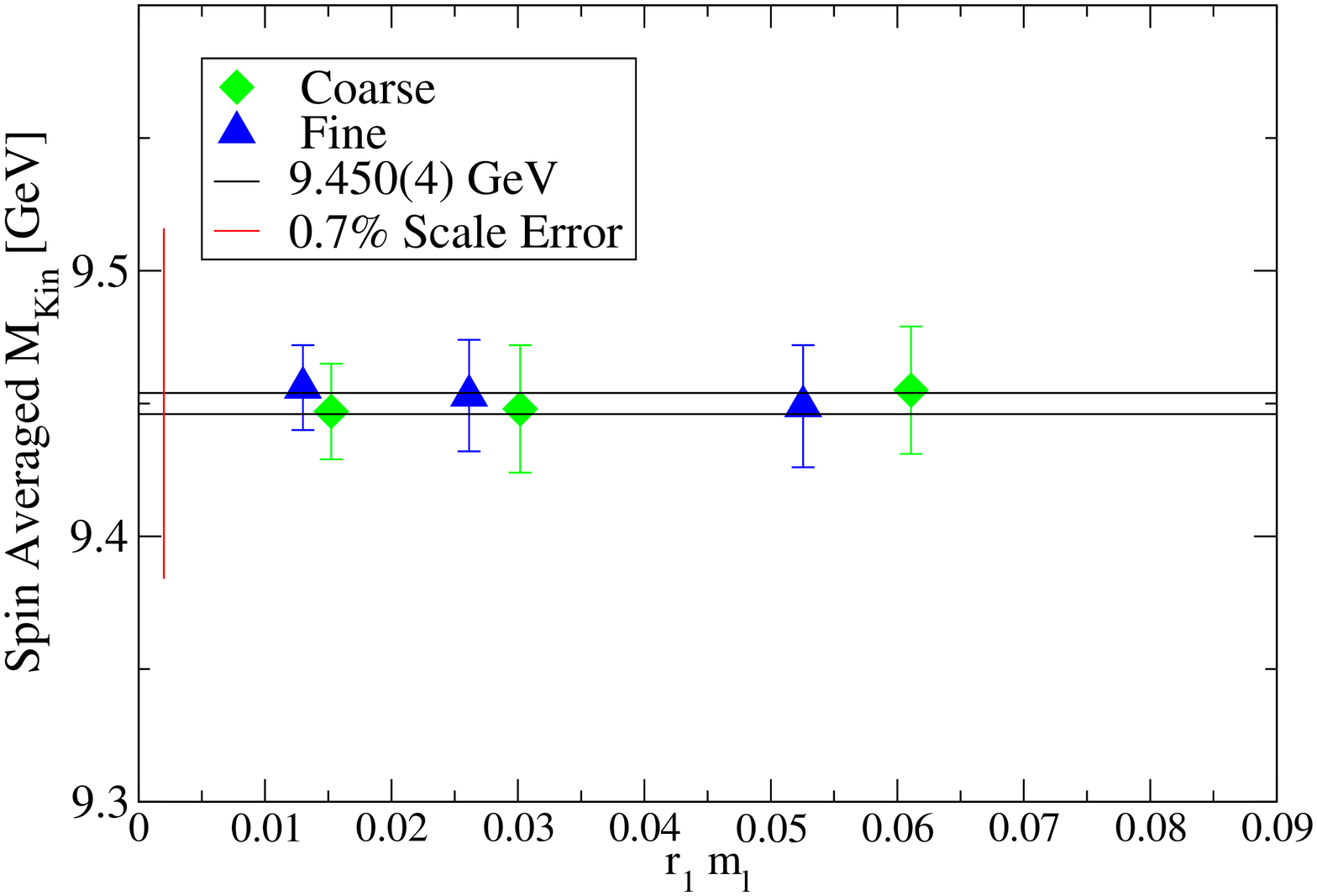}
\includegraphics[width=.45\textwidth,angle=270]{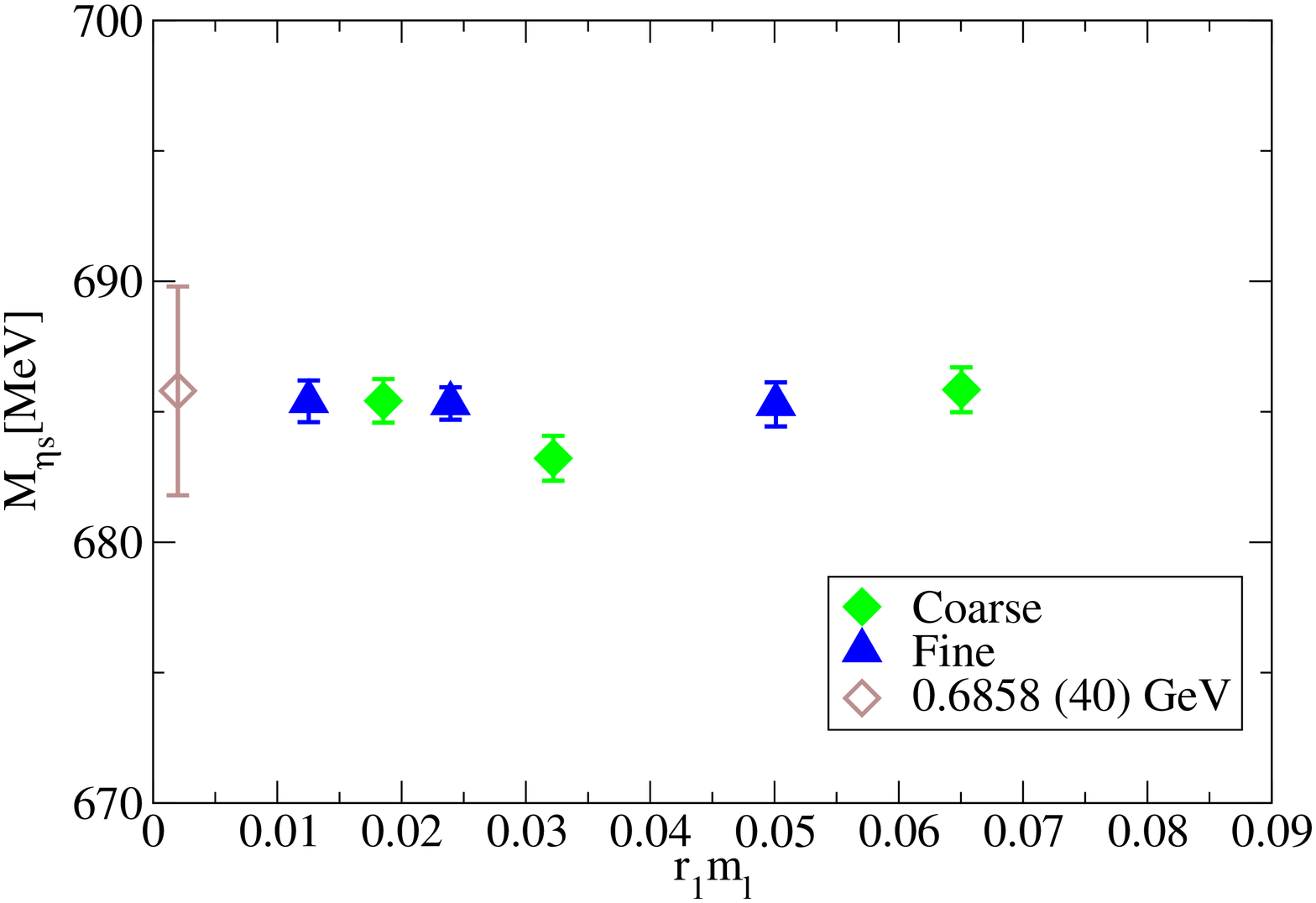}
\caption{Tuning of the $b$ (left) and $s$ (right) quark masses. }
\label{tune}
\end{figure}

Moreover, this calculation is done with the new scale parameter $r_1$ values.
HPQCD was using $r_1=0.321(5)$ fm extracted from $\Upsilon$ splittings~\cite{oldr1}.
In 2010 HPQCD published a much more accurate $r_1$ determination, $r_1=0.3133(23)$ fm, based on several physical quantities and an improved continuum extrapolation with 5 lattice spacings~\cite{newr1}.
We needed to re-tune valence quark masses due to the scale changes, and updating for such different new settings is one of the main purposes of this analysis. 
We used the spin averaged $\Upsilon$ mass to tune the bare bottom quark mass, and (fictitious) $\eta_s$ for the strange quark mass. 
Fig.~\ref{tune} shows the tuning for $b$ (left) and $s$ (right) quarks. 
As one can see, our physical target meson mass has a large error compared to the deviations of the alignment of the tuning measurements.  We found that it is important to tune quark masses precisely up to the statistical errors or $r_1/a$ errors of the tuning measurements.  
This precise tuning ensures correct estimation for $\chi^2$ of chiral and continuum extrapolation.
If one has large deviations between the tuning measurements more than its statistical errors, then the chiral and continuum extrapolations may suffer from additional systematic errors.

\begin{table}
\centering
\begin{tabular}{|c|c|c|c|c|c|}
\hline
Set &  $r_1/a$ & $m_l/m_s$ (sea)   &  $N_{conf}$&
 $N_{tsrc}$ & $L^3 \times N_t$ \\
\hline
\hline
C1  & 2.647 & 0.005/0.050   & 1200  &  2 & $24^3 \times 64$ \\
\hline
C2  & 2.618 & 0.010/0.050  & 1200   & 2 & $20^3 \times 64$ \\
\hline
C3  & 2.644 & 0.020/0.050  &  600  & 2 & $20^3 \times 64$ \\
\hline
\hline
 F0  & 3.695  &  0.0031/0.031  & 600  & 4 & $40^3 \times 96$ \\
\hline
F1  & 3.699 & 0.0062/0.031  & 1200  & 4  & $28^3 \times 96$ \\
\hline
F2  & 3.712 & 0.0124/0.031  & 600  & 4 & $28^3 \times 96$ \\
\hline
\end{tabular}
\caption{Simulation details on three ``coarse'' and three ``fine'' MILC AsqTad ensembles. $N_{conf}$ is the number of the configurations that were used in the simulation, and $N_{tsrc}$ is the number of time sources per configuration.}
\label{enb}
\end{table}

We calculated operator matching factors in full QCD at one-loop through order $\alpha_s$, $\frac{\Lambda_{QCD}}{M}$, $\frac{\alpha_s}{aM}$, $a \alpha_s$, and $\alpha_s \frac{\Lambda_{QCD}}{M}$. These matching calculations were presented separately at this conference~\cite{mat}. We also employed random-wall sources for the HISQ propagators and Gaussian smearing sources for the NRQCD propagators. 

Including all statistical and systematic errors, we obtained
\begin{equation}
f_B = 0.191(9) {\rm GeV}, \;\;\;\;\;\; f_{B_s}=0.228(10) {\rm GeV},
\end{equation}
and 
\begin{equation}
\frac{f_{B_s}}{f_B}=1.188(18).
\end{equation}
These calculations are a definite improvement on our previous calculations~\cite{fb2}, $f_B=0.190(13) {\rm GeV}$, $f_{B_s}=0.231(15) {\rm GeV}$, and $f_{B_s}/f_B=1.226(26)$.  
The largest source of errors is the operator matching error for the decay constants. 
The ratio $f_{B_s}/f_B$ has very small errors, since most of the matching factors are canceled.

If one calculates the decay constants without matching factors, one could reduce around 5 \% errors down to 1 $\sim$ 2 \% errors.  
Recently, HPQCD has calculated $f_{B_s}$ without matching factors~\cite{fbs}, and obtained $f_{B_s}=0.225(4) {\rm GeV}$ with only 1.8 \% errors. 
Essentially, this very precise calculation utilizes the HISQ action for the $b$ quark.
The HISQ action can be used to simulate the charm quark, but for the bottom quark  it is very difficult with current technology. 
The clever idea was that in fact we can simulate a heavy quark heavier than the charm quark but lighter than the bottom quark. 
Once one gets heavy meson correlators depending on multiple heavy quark masses, then one can extrapolate to the physical bottom quark mass.
In this way, one can determine the decay constants without matching factors.  

\begin{figure}
\includegraphics[width=.45\textwidth,angle=270]{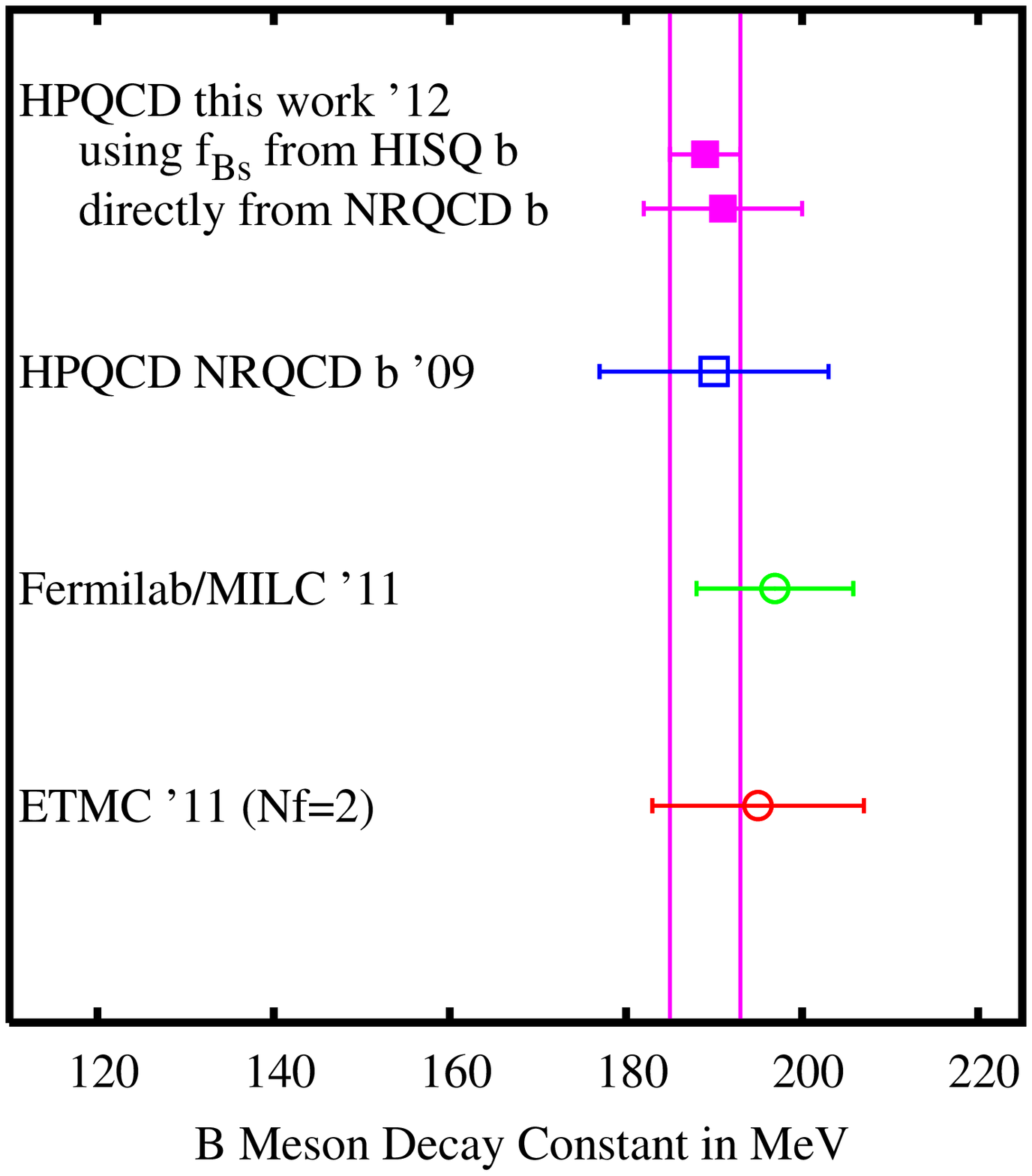}
\includegraphics[width=.45\textwidth,angle=270]{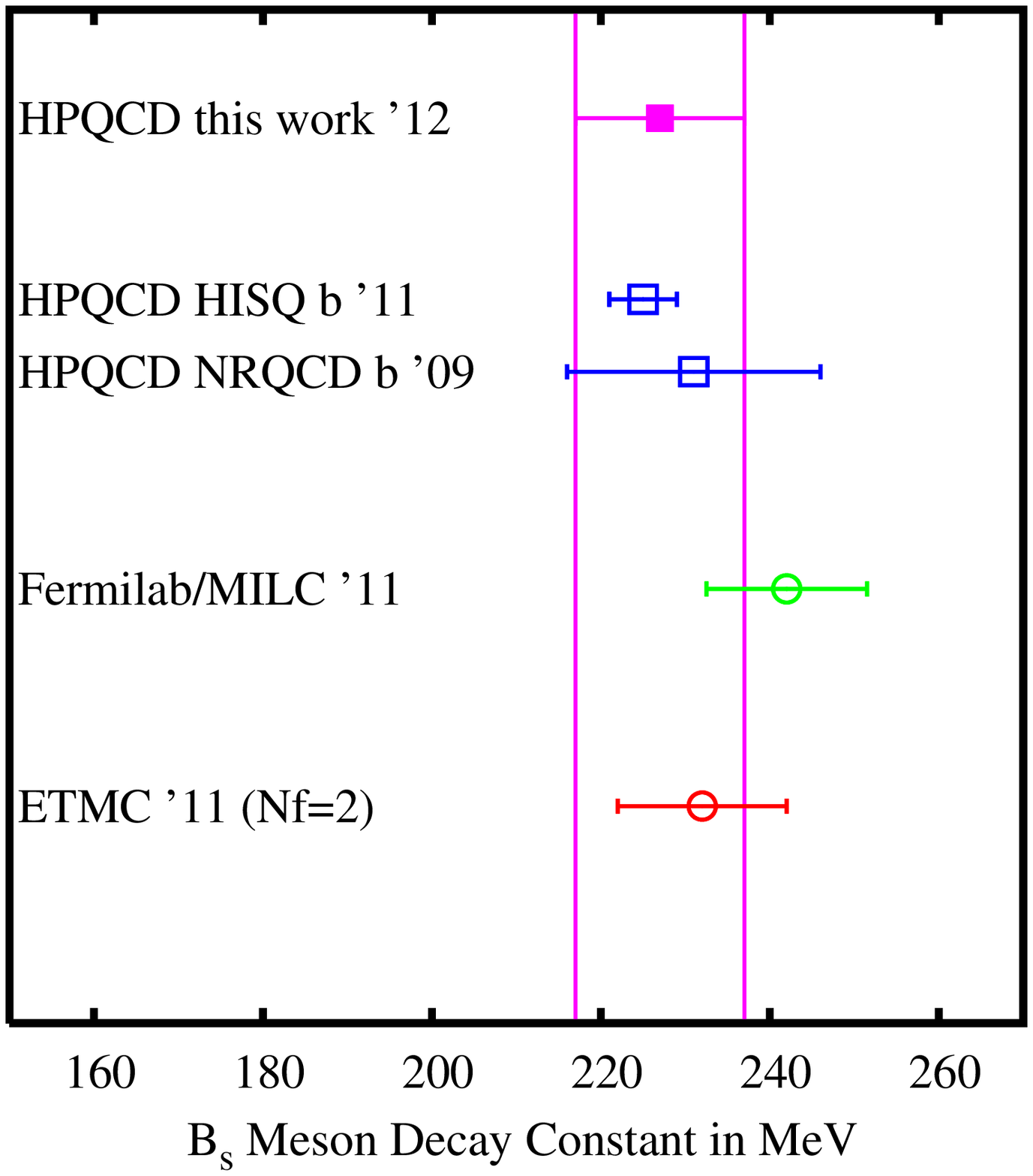}
\caption{Comparisons of results for $f_B$ (left) and $f_{B_s}$ (right) from this analysis, previous HPQCD, Fermilab/MILC and ETM collaborations.}
\label{bmeson}
\end{figure}

Of course, we can apply this heavy HISQ method for $f_B$, but it would be difficult.
First of all, the light quark is much more expensive than the strange quark, and for $f_B$ we need to perform additional chiral extrapolation. 
Thus, for now we fix $f_B$ by combining ${f_{B_s}}/{f_B}$ from the NRQCD analysis and $f_{B_s}$ from the heavy HISQ analysis;
\begin{equation}
f_B \equiv \bigg [ \frac{f_{B_s}}{f_B} \bigg]^{-1}_{NRQCD} \times f_{B_s}^{HISQ} = 0.189(4) {\rm GeV}.
\end{equation}
This $f_B$ result with 2 \% total error is the most accurate $f_B$ available today.
Comparisons of results for $f_B$ (left) and $f_{B_s}$ (right) are shown in Fig. ~\ref{bmeson}

\section{$D$ and $D_s$ meson decay constants}

With the same simulation setting shown in Tab.~\ref{enb}, we determined $D$ and $D_s$ meson decay constants.
The discretization error of the HISQ action starts at $\mathcal{O}(\alpha_s (am_h)^2 v^2/c^2)$ and $\mathcal{O}((am_h)^4 v^2/c^2)$, and this provides enough accuracy to simulate relativistic charm quarks on current typical lattices.
So, we apply the HISQ action for all valence quarks including the charm quark.
Thus, we can evaluate the decay constants without matching factors, since the HISQ action exhibits the chiral symmetry in the continuum limit. 
The decay constant can be written with the heavy-light axial vector current $A_\mu = \overline{\Psi}_q  \gamma_\mu \gamma_5 \Psi_c$, with $q=s$ or $d$, as
\begin{equation}
<0|A_\mu|D> = p_\mu f_{D_q}.
\end{equation}
We can express the decay constant in terms of the pseudoscalar density $PS=\overline{\Psi}_q \gamma_5 \Psi_c$, as we used for light meson decay constants, $f_\pi$ and $f_K$,
\begin{equation}
f_{D_q} = \frac{m_c+m_q}{M_{D_q}^2} <0|PS|D_q>.
\end{equation}

As we did for the $B$ decay constants, we also re-tuned the charm quark mass for the new $r_1=0.3133(23)$ fm. We used $\eta_c$ mass to fix the charm quark mass. 
HPQCD updated $f_{D_s}$ with the new $r_1$ in 2010~\cite{fds} already, so this work is mainly to update $f_D$ with the new $r_1$.
Our final results are
\begin{equation}
f_D = 208.3 (1.0)_{stat.}(3.3)_{sys.} {\rm MeV}, \;\;\;\;\;\; f_{D_s}= 246.0(0.7)_{stat.} (3.5)_{sys.} {\rm MeV},
\end{equation}
and,
\begin{equation}
\frac{f_{D_s}}{f_D}=1.187(4)_{stat.}(12)_{sys.},
\end{equation}
which show good agreement with our previous determinations. 

One interesting question would be what the impact of the scale change is.
Tab.~\ref{impact} summarizes HPQCD's determinations of the decay constants with the old and new scale factor $r_1$.  
We found no significant effect due to the scale change.
(One exception is for $f_{D_s}$ with the old $r_1$ and the most accurate result with the new $r_1$.)
It appears that for the decay constants re-tuning of quark masses largely compensates the shift from overall scale change. 
Thus, predicting the impact of the scale change before the actual calculations would be risky. 
We will investigate the impact of the scale change further in the future. 
This study would lead to better estimation of systematic errors for the scale setting. 

\begin{table}
\centering
\begin{tabular}{|c|c|c|}
\hline
& Old $r_1$ & New $r_1$ \\
\hline
$f_{D_s}$ & 241(3) & 246(4), 248(3) \\
$f_D$ & 207(4) & 208(3) \\
$f_{D_s}/f_D$ & 1.164(11) & 1.187(12) \\
$f_{B_s}$ & 231(15) & 228(10), 225(4) \\
$f_B$ & 190(13) & 191(9), 189(4) \\
$f_{B_s}/f_B$ & 1.226(26) & 1.188(18) \\
$f_K$ & 157(2) & 159(2) \\
$f_\pi$ & 132(2) & 132(2) \\
\hline
\end{tabular}
\caption{Decay constants from HPQCD with the old $r_1=0.321(5) $fm and the new $r_1=0.3133(23) $fm. The unit of the decay constants is MeV, and the ratios are dimensionless. }
\label{impact}
\end{table}

\begin{figure}
\includegraphics[width=.45\textwidth,angle=270]{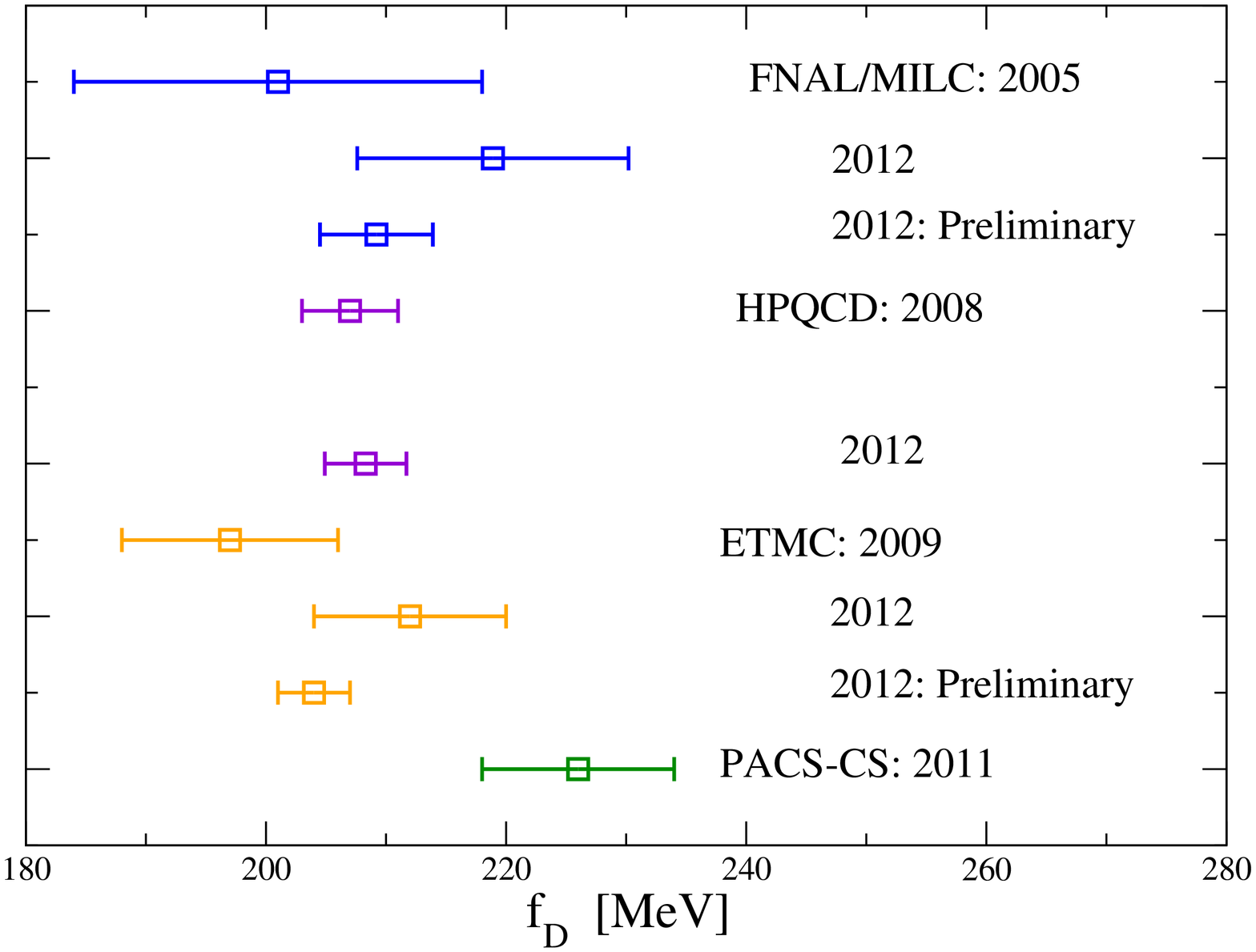}
\includegraphics[width=.45\textwidth,angle=270]{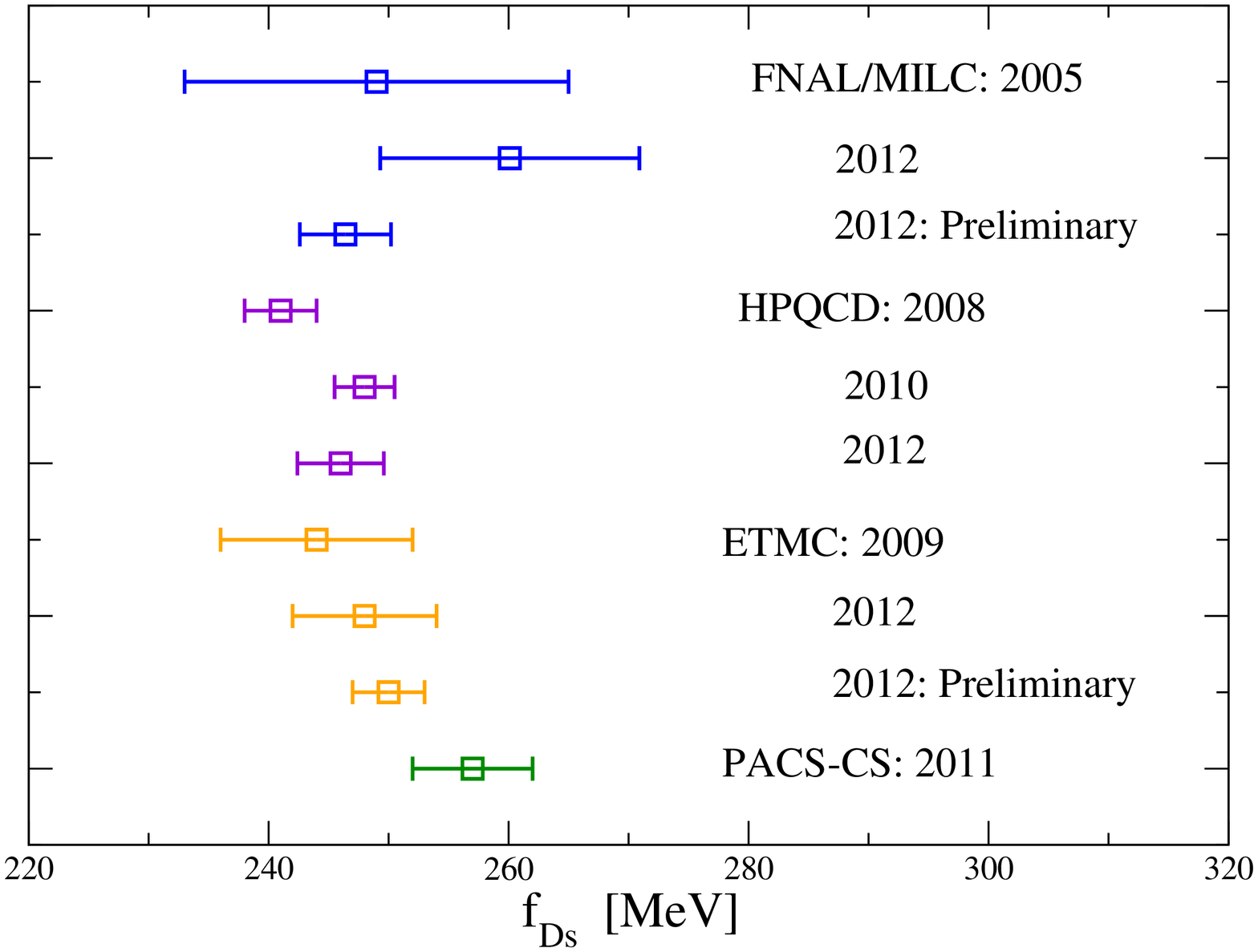}
\caption{Comparisons of results for $f_D$ (left) and $f_{D_s}$ (right). The results of this proceeding are shown under HPQCD 2012. }
\label{dmeson}
\end{figure}

In Fig.~\ref{dmeson}, we compare $f_D$ and $f_{D_s}$ results from FNAL/MILC~\cite{milc}, HPQCD~\cite{fds}\cite{fds2}, ETMC~\cite{etmc}, and PACS-CS~\cite{pacs}. The comparisons include FNAL/MILC's preliminary results with $N_f=2+1+1$ including simulations at the physical pion mass, and ETMC's preliminary results with $N_f=2+1+1$.
In their preliminary results, they achieve a good precision that is comparable to our best results, and they show very good agreement with HPQCD. 

\begin{figure}
\includegraphics[width=.45\textwidth,angle=270]{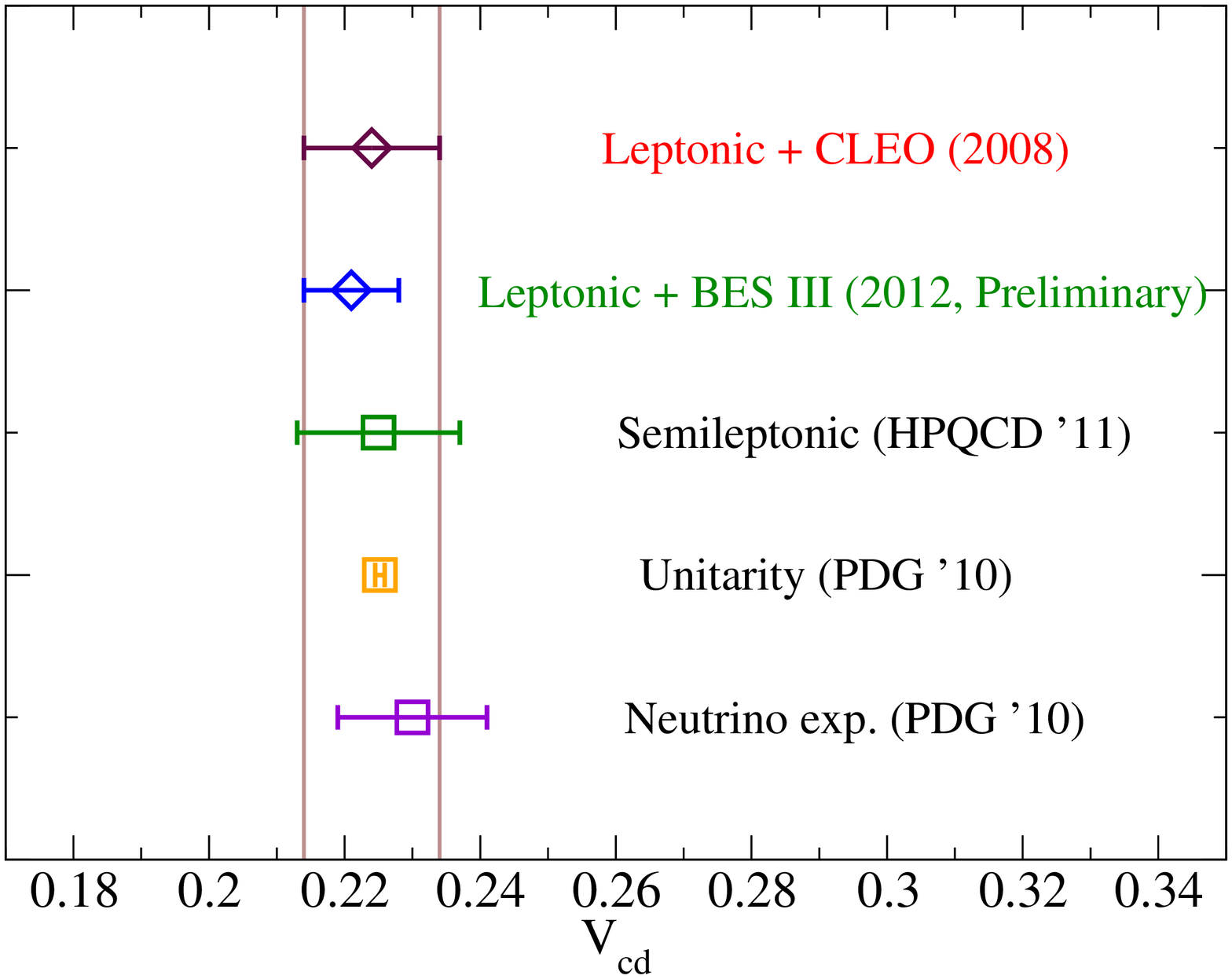}
\includegraphics[width=.45\textwidth,angle=270]{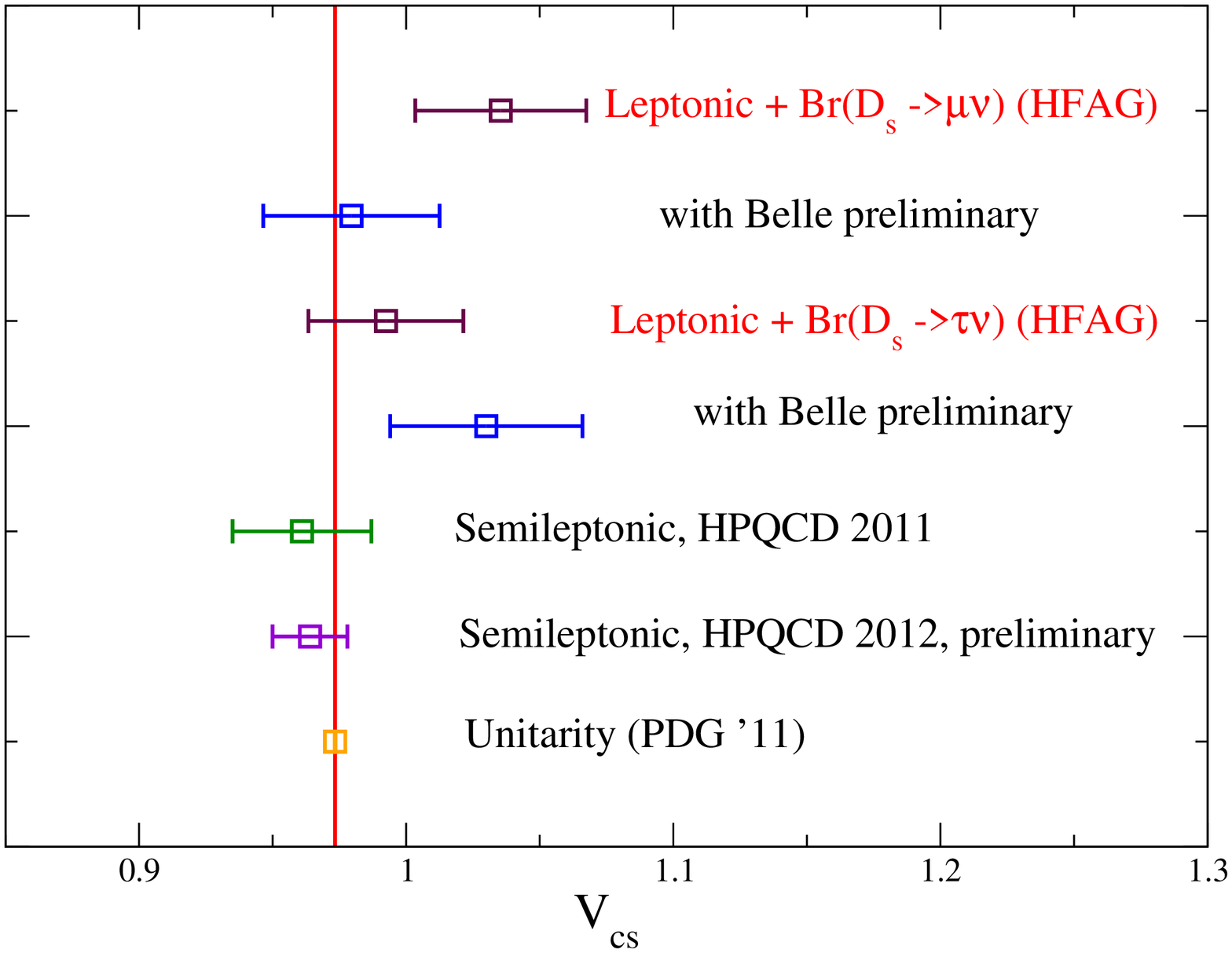}
\caption{Comparisons of results for $|V_{cd}|$ (left) and $|V_{cs}|$ (right) from leptonic decays and semileptonic decays. }
\label{ckm}
\end{figure}

Combining our decay constant results with branching fractions from experiments, we can obtain corresponding CKM matrix elements. See Fig.~\ref{ckm} for the comparisons. 
For $|V_{cd}|$, one can immediately notice that the leptonic determination and the semileptonic determination are in good agreement. Those two results were obtained with completely different systematics for both lattice and experiment analysis, since the decay channels are quite different.  
This is a highly non-trivial check for lattice formulation and experiments. 
Those two lattice determinations of $|V_{cd}|$ demonstrate good agreement with the unitarity point as well.
Thus, we do not see any signature of new physics here yet. 
We note that now the precision of the lattice determination of $|V_{cd}|$, especially with the preliminary branching fraction result from BES III is actually better than the accuracy of the determination from neutrino experiments. So far, the PDG quotes the neutrino experiment result for  $|V_{cd}|$. This is simply because lattice determinations had much larger errors in the past. 

For $|V_{cs}|$, the situation is more interesting. The right plot of Fig.~\ref{ckm} shows that the semileptonic determination and the unitarity point are in good agreement. 
However, for the leptonic determinations, it shows some discrepancies depending on the decay channels and experiments. 
If we only consider average leptonic determinations with averages of HFAG, then the $|V_{cs}|$ represents more than 1 $\sigma$ discrepancy between the unitarity point, which is the identical to observation of the $f_{D_s}$ puzzle.  
The $f_{D_s}$ puzzle~\cite{puzzle} was a 4 $\sigma$ tension between experiments and lattice determinations of $f_{D_s}$ in around 2010. 
The puzzle is no longer a puzzle, since new experiments and lattice analysis results have moved closer to each other. Now the difference is about 1.6 $\sigma$. 
When experiments determine $f_{D_s}$, they use the unitarity $|V_{cs}|$. 
Thus, the $f_{D_s}$ puzzle indicates a difference between $|V_{cs}|$ from the unitarity point and our determination from the leptonic decay of $D_s$ meson. 
In Fig.~\ref{ckm}, we show the results with two Belle's 2012 preliminary results, the first is for $D_s \rightarrow \mu \nu$ and the second is for $D_s \rightarrow \tau \nu$; these results suggest that we need to wait to see the experiments attain more accuracy.
It may still be possible to see some discrepancy in determinations of $|V_{cs}|$, and this could be a hint for new physics.

\acknowledgments
This work was supported by the DOE  and
the NSF in the U.S., by the STFC in the U.K., 
by MICINN and DGIID-DGA in Spain, and by ITN-STRONGnet in the EU.  
Numerical simulations were
 carried out on facilities
of the USQCD collaboration funded by the Office of Science of the DOE and 
at the Ohio Supercomputer Center.
We thank the MILC collaboration for use of their gauge configurations.

\end{document}